\documentclass[12pt, twoside, twocolumn, a4paper]{article}
\usepackage{xfrac}
\usepackage{abstract}
\usepackage[utf8]{inputenc}
\usepackage[T1]{fontenc}
\usepackage{mathptmx}
\usepackage[fleqn]{amsmath}
\usepackage{geometry}
\usepackage{mwe}
\usepackage{afterpage}

\usepackage{graphicx}
\usepackage[export]{adjustbox}

\usepackage{caption}
\usepackage{enumerate}
\usepackage{hyperref}

\usepackage{authblk}

\usepackage{color}

\usepackage{fancyhdr}
\newcommand{\nine}{\fontsize{9pt}{\baselineskip}\selectfont}    
\newcommand{\ten}{\fontsize{10pt}{\baselineskip}\selectfont}    

\usepackage{titlesec}
\titleformat{\section}{\fontsize{16}{0}\selectfont}{\thesection}{1em}{\normalfont}
\titleformat{\subsection}{\fontsize{14}{0}\selectfont}{\thesubsection}{1em}{\itshape}
\titleformat{\subsubsection}{\fontsize{12}{0}\selectfont}{\thesubsubsection}{1em}{\itshape}

\usepackage{amsmath}

\usepackage{mystyle}
\usepackage{bbm}
\hypersetup{colorlinks=true,citecolor=blue,urlcolor=blue,linkcolor=black}

\fancypagestyle{titlepagestyle}{%
\fancyhf{}
\fancyfoot[L]{\nine
Proceedings of the 13th International Conference on Structural 
Safety and Reliability (ICOSSAR 2021-2022), China.
}

}

\begin{document}

\title{Accelerated Weight Histogram Method\\for Rare Event Simulations}

\author[a]{Jack Lidmar}
\author[b]{Johan Spross}
\author[c]{John Leander}

\affil[a]{Department of Physics, KTH Royal Institute of Technology, SE-106 91 Stockholm, Sweden}
\affil[b]{Soil and Rock Mechanics, KTH Royal Institute of Technology, SE-100 44 Stockholm, Sweden}
\affil[c]{Structural Engineering and Bridges, KTH Royal Institute of Technology, SE-100 44 Stockholm, Sweden}

\twocolumn[
\date{\vspace*{-1em}\ten Submitted: October 15, 2021; Accepted: May 7, 2022}
\maketitle
\thispagestyle{titlepagestyle}
\vspace*{-2em}

\begin{abstract}
	\noindent
        ABSTRACT: We describe an adaptive Markov chain Monte Carlo method suitable for the estimation of rare failure probabilities in complex probabilistic models. This method, the Accelerated Weight Histogram (AWH) method, has its origin in statistical physics~\citep{AWH} and has successfully been applied to molecular dynamics simulations in biophysics.  Here we introduce it in the context of structural reliability and demonstrate its usefulness for calculation of failure probabilities in some selected problems of varying degrees of complexity and compare with other established techniques, e.g., subset simulations.
\end{abstract}

\vspace*{20pt}
]


\section{Introduction}
\vspace{-1em}
Estimation of very small probabilities is of interest in many science and engineering fields, which has spurred the development of different methods for this challenging task. For structural safety assessment, available methods range from approximate methods like first-order second-moment methods \citep[see e.g.][]{Hasofer1974,Rackwitz1978}, to more advanced and more precise methods, which are often sampling-based like importance sampling \citep[e.g.][]{Au1999,Papaioannou2016} and subset simulation \citep{Au2001,Au2014}. Following the development of more complex probabilistic models, higher demands are put on the efficiency of the probability estimation methods. Though, as it turns out, a given probability estimation method can be more or less suitable to the probabilistic model at hand; \citet{Straub2016} illustrated this clearly. Developing new -- or improving old -- probability estimation methods for structural reliability problems is therefore an important task for research.

In this contribution we present a novel adaption of the simulation-based Accelerated Weight Histogram (AWH) method to estimations of rare failure probabilities in structural reliability models. The AWH method was developed by 
\citet{AWH}, originally for tackling difficult-to-sample problems in statistical and biological physics. The generality of the AWH framework makes it applicable to the simulation of general probabilistic models and, in particular, to the calculation of rare event probabilities. 

\begin{figure}
\centering
\includegraphics[width=0.25\textwidth]{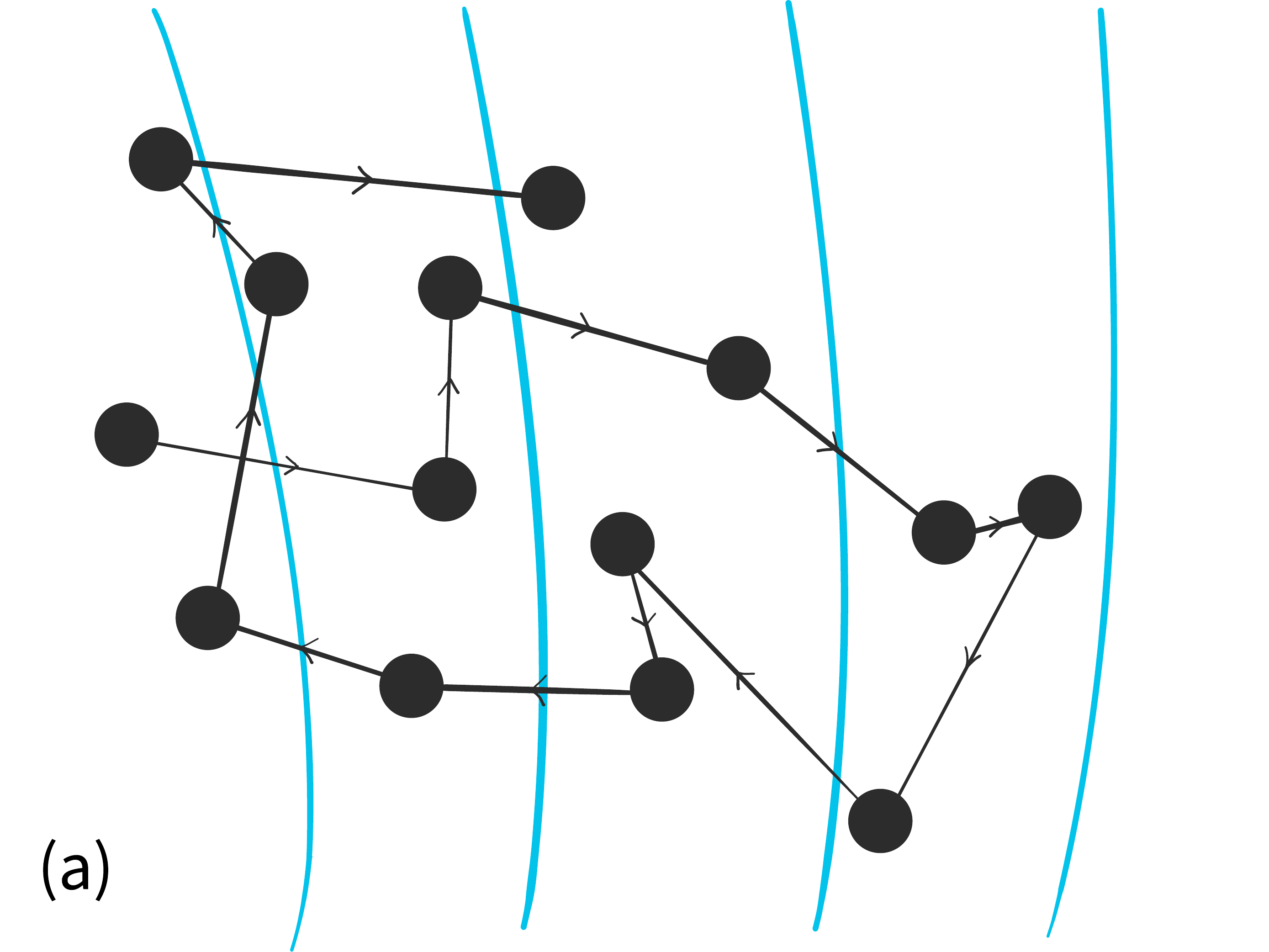}%
\includegraphics[width=0.25\textwidth]{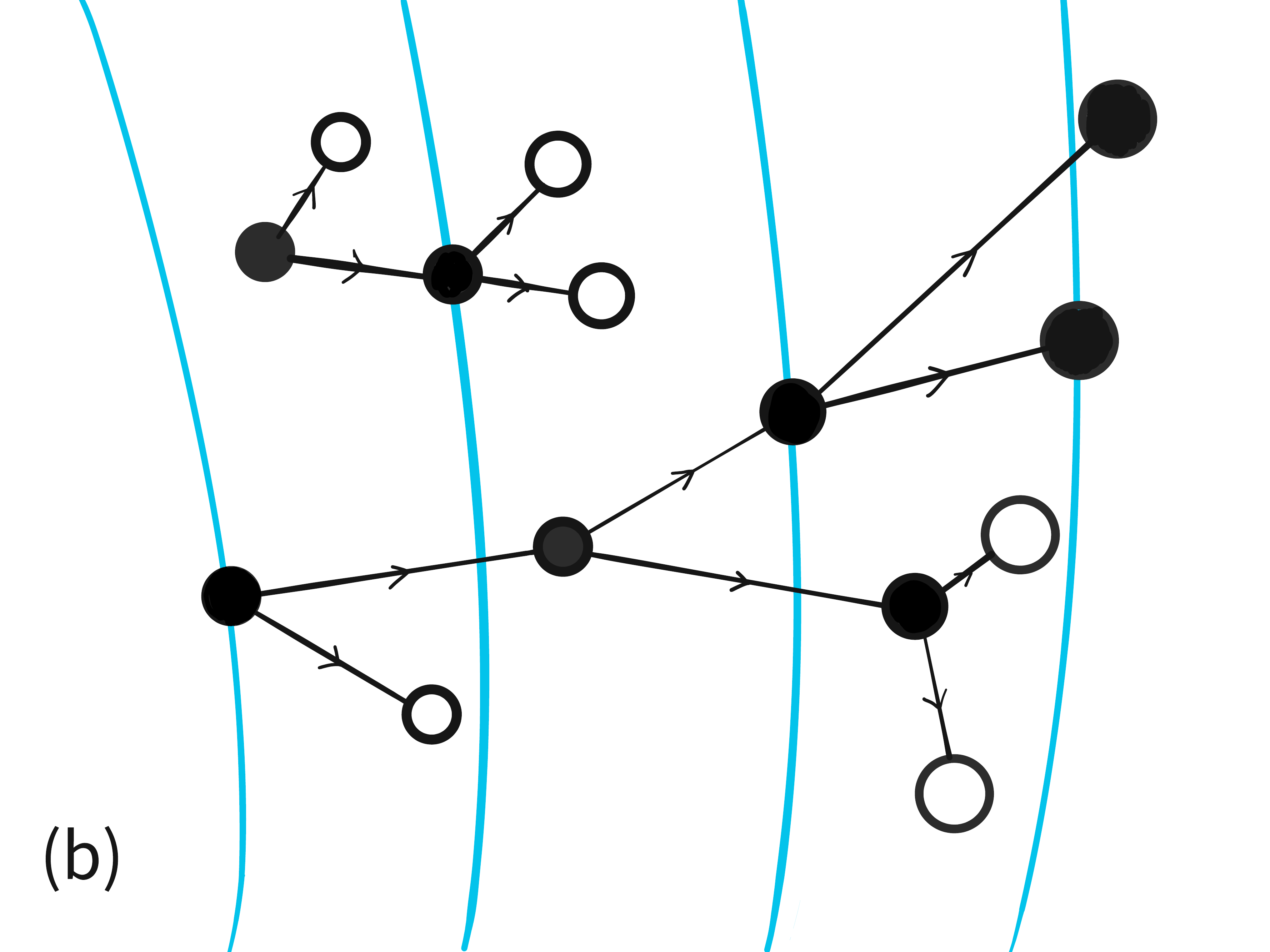}
\caption{\label{fig:illustration}%
Figure 1. Illustration of the difference between (a) AWH and (b) subset simulations.
The blue lines delimit the different subsets $\F_k$.}
\end{figure}

Similarly to subset simulations, a sequence of level subsets
$\F \equiv \F_0 \subset \F_1 \subset \cdots \subset \F_M \equiv \Omega$
is introduced, where $\F$ denotes the rare event whose probability is to be estimated and
$\Omega$ the set of all events.
The failure probability may then be expressed as $\pi(\F) = P(\F_0 | \F_1) P(\F_1|\F_2) \cdots P(\F_{M-1}|\F_M)$.
In subset simulations the conditional failure probabilities $P(\F_k | \F_{k+1})$ for increasingly rare events are estimated starting from a large population of independent samples generated from the original unrestricted distribution $P(x | \F_M) \equiv \pi(x)$, and then moving one-way towards the rarer regions using a combination of Markov Chain Monte Carlo updates and splitting to generate samples whose marginal distribution is $P(x|\F_k)$ at each level, see \Fig{fig:illustration}(b).
In contrast, the AWH method uses adaptive importance sampling by designing a Markov chain that carries out a guided random walk among the levels as illustrated in \Fig{fig:illustration}(a). The resulting stochastic process thus jumps back and forth many times among the different levels.


This paper first describes the AWH method, as adapted to rare event simulations. The method is then applied to two probabilistic model examples and its efficiency is compared against the subset simulation method, which today gains increasing popularity for structural reliability assessments.

\section{The accelerated weight histogram (AWH)  method}

The AWH method~\citep{AWH} works in an extended configuration space where one or more parameters in a probabilistic model are promoted to dynamical variables.  By randomly transitioning between different parameter values the sampling of complicated distributions is often enhanced, for example in rarely visited regions.
First we describe the idea in a completely general setting, and then specialize to the case of rare events characterized by a limit state function.

Let $x \in \Omega$ denote a vector of stochastic variables (often high-dimensional), whose probabilities depend on a parameter vector $\lambda$, i.e.\ $x \sim P(x|\lambda)$.
It is convenient to work with logarithms and write the conditional distribution as
\begin{equation}												\label{eq:cond}
P(x|\lambda) = e^{F_\lambda - E_\lambda(x)},
\end{equation}
where often the normalization factors $e^{F_\lambda}$ are unknown from the start.
On the other hand we assume, as is usually always the case, that we can generate samples from $P(x|\lambda)$, via e.g.\ Markov chain Monte Carlo (MCMC) without knowing $F_\lambda$.
A joint distribution of $x$ and $\lambda$ is defined as
\begin{equation}												\label{eq:joint}
P(x, \lambda)
= \f 1 \Z e^{f_\lambda - E_\lambda(x)},
\end{equation}
which depends on a new set of tuning parameters $f_\lambda$ that may be used to control how the samples will be distributed among the different $\lambda$.
In fact, the marginal distribution of $\lambda$ is given by
\begin{equation}												\label{eq:marginal}
P(\lambda) = \int_\Omega P(x,\lambda) dx = \f 1 \Z e^{f_\lambda - F_\lambda},
\end{equation}
where $\Z = \sum_\lambda e^{f_\lambda - F_\lambda}$.
By tuning the $f_\lambda \approx F_\lambda + \ln \pi_\lambda$ it is possible to make the marginal distribution approach any desired target distribution $\pi_\lambda$.
This is quite nontrivial since $F_\lambda$ is typically unknown,
but the AWH framework provides an efficient way to accomplish this.
In AWH a Markov chain with a joint equilibrium distribution given by \Eq{eq:joint} is constructed by alternating between MCMC moves that leave either $P(x|\lambda)$ or $P(\lambda|x)$ invariant,
while adaptively fine tuning the hyper parameters $f_\lambda$~\citep{AWH}.
Often the target distribution $\pi_\lambda$ is chosen to be uniform, $\pi_\lambda=$ const., in the given parameterization.
While this typically works well, this is really an arbitrary choice, and it is sometimes possible to improve the efficiency by optimizing $\pi_\lambda$ (see the discussion in Sec.~\ref{sec:target}).

\subsection{Rare events}

Suppose that we are interested in estimating the probability of some rare failure event,
$\pi(x \in \F)$.
Here $\pi(x)$ is a probabilistic model of the system of interest, and $\F \subset \Omega$ the set of events corresponding to failure of the system.
It is often convenient to describe this set using a limit state function $G(x)$, so that $\F = \{x | G(x) \leq 0 \}$, and the sought probability is
$\pi(\F) = \pi(G(x) \leq 0)$.
If this probability is very small it is hard to get an accurate estimate of it using a crude MC estimator,
\begin{equation}												\label{eq:crude}
\pi(\F) \approx \frac{1}{N} \sum_{n=1}^N \mathbbm{1}(G(x_n) \leq 0) ,
\end{equation}
where the $x_n$ are samples from $\pi(x)$ and $\mathbbm{1}(\cdot)$ is the indicator function, which is 1 if the argument is true, 0 if false.
The variance of \Eq{eq:crude} is $\pi(\F)(1-\pi(\F)) / N$, so that a sample size $N \gg 1/\pi(\F)$ is needed for a small relative error.
A further complication is that $G(\cdot)$ might be very expensive to evaluate, e.g., it might be the result from a finite element calculation or so.

\subsection{Joint distribution for rare events}

In order to put the problem of estimation of rare event probabilities on a form suitable for the AWH method we need to introduce some parameter $\lambda$,
which interpolates between $\pi(x)$ and $P(x | \F)$.
To this end, let us introduce the sequence of distributions
\begin{align}												\label{eq:restricted}
P(x | \lambda_m ) &= e^{F_m} \mathbbm{1}(G(x) \leq \lambda_m) \pi(x),
\\ \nonumber
m &= 0, \ldots, M
\end{align}
where $\{ 0 = \lambda_0 < \lambda_1 < \ldots < \lambda_M = \infty\}$ form a monotonically increasing sequence of threshold levels.
The normalization factors are given by
\begin{align}												\label{eq:normali}
e^{-F_m} &= \int_{\Omega} dx \mathbbm{1}(G(x) \leq \lambda_m) \pi(x)
\nonumber \\
&= \pi(G \leq \lambda_m) .
\end{align}
Thus, the probability of the failure event is $\pi(G \leq 0) = \pi(G \leq \lambda_0) = e^{-F_0}$,
while $\pi(G \leq \lambda_M) = \pi(G \leq \infty) = e^{-F_M} = 1$.
We now define a joint distribution of $x$ and $\lambda_m$ as
\begin{equation}												\label{eq:joint-rare-event}
P(x , \lambda_m ) = \f 1 \Z e^{f_m} \mathbbm{1}(G(x) \leq \lambda_m) \pi(x) .
\end{equation}
The marginal probability of $\lambda_m$ or equivalently the level index $m$ becomes
\begin{equation}												\label{eq:marginal-rare-event}
P_m = \f 1 \Z e^{f_m - F_m},
\qquad
\Z = \sum_{k=0}^{M} e^{f_k - F_k}.
\end{equation}
To make this approach the desired target distribution $\pi_m$ one needs to fine tune $f_m \approx F_m + \ln \pi_m$.
When converged, the AWH algorithm will produce samples from \Eq{eq:joint-rare-event} and estimates of the logarithmic normalization constants $F_m$ (up to a common additive constant).
The sought rare event probability may then be calculated as
\begin{equation}                \label{eq:Pf}
\pi(\F) = \pi(G(x) \leq 0) = e^{F_M - F_0}.
\end{equation}

%
%

\subsection{The AWH algorithm for rare events}

The AWH algorithm generates samples from the joint distribution \Eq{eq:joint-rare-event} by interleaving MCMC moves $x \to x'$ at fixed $\lambda_m$ and moves $m \to m'$ at fixed $x$.
For the former we can use any MC method that leaves $P(x|m)$ invariant.
For the latter it is convenient to use a Gibbs sampler, drawing $m$ from the conditional distribution
\begin{equation}												\label{eq:m-given-x}
P(m|x) \equiv w_m(x) = \frac{e^{f_m} \mathbbm{1}(G(x) \leq \lambda_m)}
{\sum_k e^{f_k} \mathbbm{1}(G(x) \leq \lambda_k)} .
\end{equation}
During the simulation, a histogram of weights $W_k = \sum_t w_{k}(x_t)$ is accumulated and used to update the hyper parameters $f_k$.
The whole algorithm is summarized as follows:
\\[1em]
Initialize $m = M$ and $x \sim P(x|\lambda_M) \equiv \pi(x)$.\\
Repeat for $n=1, ..., N_\mathrm{it}$ (or until the desired accuracy has been reached):
\begin{enumerate}
\item
Carry out one or more MCMC steps at fixed $m$:
\begin{enumerate}
\item Propose a new state $x'$ with probability $q(x'|x)$.
\item Accept, i.e.\ set $x \gets x'$, if $q(x|x') \pi(x')/q(x'|x)\pi(x) \geq u$ and
	$G(x') \leq \lambda_m$,
	where $u \sim U[0,1)$ is a uniform random variate in $[0,1)$.
\item \label{alg:reject}
	Otherwise set the new state equal to the old one, $x \gets x$.
\end{enumerate}
\item
Calculate the weights $w_k(x) \equiv P(k|x)$ for all $k$, using \Eq{eq:m-given-x}, and update the weight histogram, $W^{(n)}_k = W^{(n-1)}_k + w_k(x)$.
\item
Choose a new level index $m$ with probability $w_m(x)$.
\item
Update the hyper parameters:\\
$f^{(n)}_k = f^{(n-1)}_k + \Delta f_k$, \ for all $k$,
where
\begin{equation}				\label{eq:f-update}
\Delta f_k = -\ln \left( \frac{W^{(n)}_k}{W_k^{(n-1)} + \pi_k} \right) .
\end{equation}


\end{enumerate}
An estimate of the failure probability is then obtained from \Eq{eq:Pf}, using $F_k \approx f_k - \ln \pi_k$.

As already mentioned, step 1 can be replaced by any MCMC update leaving the conditional $P(x|\lambda)$ invariant.
For instance one can use a Gibbs sampler instead.
When $\pi(x)$ is a multidimensional standardized normal distribution one may, e.g., use the procedure suggested in \citet{Papaioannou2015} or \citet{Au2016},
drawing each proposal $x'_i$ from
$q(\cdot | x) = N(\sqrt{1 - s^2} x_i, s^2)$, where $0<s \leq 1$
is a suitably chosen step length, and accepting if $G(x') \leq \lambda_m$.
Each time the level $\lambda = \infty$ (i.e.\ $m = M$) is visited it is obviously better to sample $x$ directly from the base probability $\pi(x)$.

Before the algorithm starts the hyper parameters $f_k$ must be initialized with a first guess (e.g., $f_k^{(0)} = 0$ if nothing is known), and the weight histogram $W_k^{(0)} = N_\mathrm{init} \pi_k$, where $N_\mathrm{init}$ is a small number, typically of order $1$ or $M$, which quantifies our prior belief in the initial guess.
As the simulation goes on the weight histogram will converge towards $N \pi_k$ and the updates will become smaller and smaller $\Delta f_k \sim 1/N \to 0$, while $f_k \to F_k + \ln \pi_k$ when the total number of (effective) samples $N$ grows large.
In the early stages of the algorithm the estimates $f_k$ will, however, typically be very poor, which can be diagnosed by a very skewed weight histogram $|W_k - N \pi_k| \gg 0$.
When this happens it is recommended to reduce the effective number of samples $N = \sum_k W_k$,
e.g.\ by setting $W_k \gets \min(W_k, c N \pi_k)$ for all $k$ and $N \gets \sum_k W_k$, where $c$ is a suitable relative tolerance for the acceptable deviations.
This prevents large peaks from building up in the weight histogram.
A value of $1.25 \leq c \leq 2$ seems to work well in many cases.

\subsection{Target distribution}
\label{sec:target}
Two related parameter choices affect the efficiency of the simulation algorithm,
namely the spacing $\delta\lambda_m$ of the levels $\lambda_m$ and the target distribution $\pi_m$.
One of the advantages of the AWH method is the ability to make large jumps along the $\lambda$-coordinate by the use of a Gibbs sampler [\Eq{eq:m-given-x}].  This makes it possible to choose a rather dense set of levels without loosing efficiency.
The distribution of the levels $\lambda_m$ or equivalently the target distribution $\pi_m$, since what matters is the product $\pi_m \delta \lambda_m$, is something which may be optimized for convergence, although this is a nontrivial problem.
In subset simulations~\citep{Au2001} the levels $\lambda_m$ may elegantly be chosen adaptively, such that $\pi(G\leq\lambda_m)/\pi(G\leq\lambda_{m+1}) = p_0$ is constant with a preset value $p_0 = 0.1 - 0.5$.
In AWH it is easier to keep the levels fixed and instead adapt the target probabilities $\pi_m$.
The simplest and most common choice would be a uniform distribution $\pi_k = 1/(M+1) = \text{const.}$, with a uniform level distribution $\lambda_k = k \,\delta \lambda$, $k=0,\ldots, M-1$ and $\lambda_M = \infty$.
The largest finite level $\lambda_{M-1}$ can be set to the largest value of $G(x)$ encountered in a short initial unrestricted MC simulation.
Instead of using a uniform distribution, it is possible to mimic the distribution implicitly used in subset simulations by setting
$\pi(\lambda) \propto |dF/d\lambda|$.
A clear benefit of this choice is that it is invariant under nonlinear reparameterizations
$\lambda \mapsto g(\lambda)$.

In practice we have to work with its discrete approximation $\pi_k \propto |\delta F_k|$, e.g.\ using central differences
[$\delta F_k = \half \left( F_{k+1} - F_{k-1} \right)$,
for $0 < k < M$ and $\delta F_0 = F_1 - F_0$, $\delta F_M = F_M - F_{M-1}$],
and with estimates of the logarithmic normalization constants $F_k \approx f_k - \ln \pi_k$ instead of the exact ones.
Since these estimates will be unreliable at the early stages of the simulation it is better to mix in also a (small) uniform component, setting
\begin{equation}								\label{eq:nonuniform}
\pi_k = \alpha \frac{1}{M+1} + (1-\alpha) \dfrac{1}{Z} \left| \delta F_{k} \right|,
\end{equation}
where $Z$ is a normalization constant and $0 \leq \alpha \leq 1$ decreasing as the simulation goes on.
In fact it seems helpful to always keep a small uniform component to increase the robustness of the algorithm.
For example, we may set $\alpha = \gamma/(\gamma + \min W_k) + \epsilon$, with $\gamma \simeq 10$ to $200$ and $\epsilon = 0.01$.
The adjustment of the target distribution may be carried out between step 1 and 2 in the algorithm.



\section{Examples}

\subsection{Normal distribution}
\label{sec:normal}

\begin{figure}[t]
	\centerline{\includegraphics[width=0.5\textwidth]{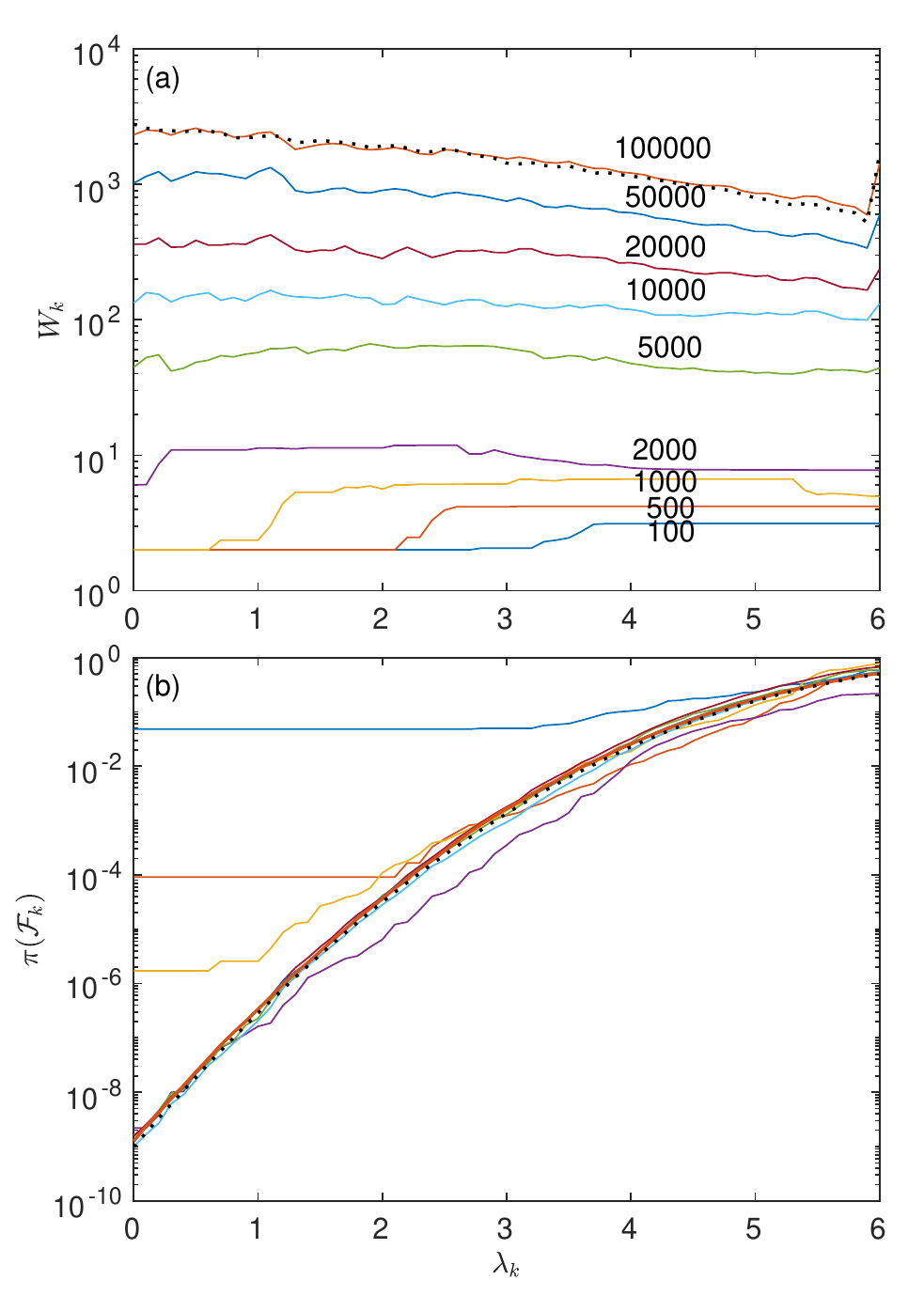}}
    \caption{\label{fig:W}Figure 2.
		 (a) Weight histogram and (b) estimate of $\pi{(G \leq \lambda_k})$ after 100, 500, 1000, 2000, 5000, 10000, 20000, 50000, and 100000 iterations of an AWH simulation of the normal distribution example (Sec.~\ref{sec:normal}). The dotted line in (a) shows the target distribution $N \pi_k$ and in (b) the exact solution $\Phi(\lambda - 6)$.
	}
\end{figure}

A simple test case is to take as base probability a multidimensional normal distribution, $x_i \sim \pi(\cdot) = N(0,1)$, and define the failure region as
$\F = \{x | G(x) \leq 0 \}$ using a limit state function $G(\mathbf x) = \beta - n^{-1/2} \sum_{i=1}^{n} x_i$.
In this case, since the sum of normal random variables is again normal, the exact failure probability is $\pi(\F) = \int_\beta^\infty \pi(x) dx = \Phi(-\beta)$,
where $\Phi(\cdot)$ is the standard normal cumulative distribution function.
For concreteness we set $n=2$ and $\beta = 6$,
which gives
$\pi(\F) \approx 0.9866 \cdot 10^{-9} \approx 10^{-9}$.

For the AWH simulations
we set the intermediate levels to $\lambda_k = 0.1 k$, for $k=0,1, \ldots, 60$, $\lambda_M = \infty$, $M=61$,
and use \Eq{eq:nonuniform} as the target distribution.
The MCMC updates of $x_i$ use a proposal $q(\cdot|x_i) = N(\sqrt{1-s^2} x_i,s^2)$, $s = 0.5$.
In order to illustrate the convergence we show in
Figure~\ref{fig:W}(a) the accumulated weight histogram $W_k$ after
$N = 100$, 500, 1000, 2000, 5000, 10000, 20000, 50000, and $100000$ iterations.
In (b) the corresponding estimates of $\pi(G \leq \lambda_m)$ are shown together with the exact solution (dotted line).
The resulting final estimate gave $\pi(\F) \approx 1.25 \cdot 10^{-9}$ after 100000 iterations.

In order to asses the error we repeated the simulations 50 times and calculated the root-mean-square (RMS) deviation from the exact result.
This gave a relative error of 0.2 after 100000 iterations (and equally many evaluations of $G(x)$) for the AWH simulations.
For comparison we also ran 50 subset simulations using a population of 10000 systems and 1000 seeds (corresponding to a conditional probability $p_0 = 0.1$ between subsequent levels).  The RMS error for the subset simulations was 0.17 using 100000 evaluations of $G(x)$.
Thus, for this test case the efficiency of AWH and subset are comparable, giving a slight edge to the subset simulation method.


\subsection{The Fiber Bundle Model (FBM)}

Here we consider another, more challenging example, the Fiber Bundle Model~\citep{Peirce1926,Daniels1945,Pradhan2010}, which consists of a load connected by $N$
parallel elastic strings or fibers.
The model was introduced to describe the strength of textiles, but variants have found widespread use in a variety of different contexts, such as fracture in materials, wire cables, earthquakes, and landslides~\citep[see e.g.][]{Pradhan2010,Faber2003,Sornette1992,Cohen2009}.

The strings have identical spring constants $\kappa$, but break at different random strains $x_i$, which are independent and identically distributed according to $\pi(x_i)$.
When the load is slowly increased from zero to a final value $L$, the weaker springs will break and the stress will be redistributed among the surviving ones. As a result the increased stress can cause additional failures and so on.
In fact, for large $N$ the failure of the structure occurs via a series of avalanches or bursts, which follow a powerlaw distribution on approaching the critical load $L_c$ at which all strings have snapped.
The total force as function of the extension $\epsilon$ is
\begin{equation}
F(\epsilon) = \sum_{i=1}^{N} \kappa \epsilon \theta(x_i-\epsilon),
\end{equation}
where the Heaviside function is defined as $\theta(x) = 0$ if $x<0$, and $1$ if $x \geq 0$.
The effect of the Heaviside function is to include only the strings with extensions less than their threshold, $\epsilon \leq x_i$.
On average the force becomes
$\overline{F}(\epsilon) = \mathbbm{E}[F(\epsilon)] = N\kappa \epsilon (1-\Pi(\epsilon))$, where $\Pi(\cdot)$ is the cumulative distribution of $x_i$.
The variance is $\mathrm{Var}\ F(\epsilon) = N \kappa^2 \epsilon^2 \Pi(\epsilon) (1-\Pi(\epsilon))$.
In the limit $N \to \infty$ the relative fluctuations around the mean tend to zero so that the structure almost surely holds for $L < L_c$ and fails for $L > L_c$, where $L_c = \max\limits_\epsilon \overline{F}(\epsilon)$.
For concreteness we set $\kappa = 1$ and assume that the thresholds are uniformly distributed, say between 0 and 1. Then $\Pi(x) = x$ and one finds a maximum force $L_c = N/4$ corresponding to an extension $\epsilon_c = 1/2$.

We now turn to the numerical estimation of the failure probability when the applied load is below the threshold $L_c$, which can occur as a result of rare fluctuations for finite $N$.
The limit state function is
\begin{align}
G(\mathbf x) &= \max_\epsilon F(\epsilon) - L
\nonumber \\
&= \kappa \max_j x_j \sum_{i=1}^N \theta(x_i - x_j) - L .
\end{align}
As an example we set $N = 1000$, $L = 200 < 250 = L_c$ and
study the case where the thresholds are uniformly distributed, $x_i \sim U[0,1)$.
For the MCMC updates we use simple Metropolis attempts in which the threshold $x_i$ of a randomly selected string $i$ is replaced by a uniform random variate in $[0,1)$.
In the AWH simulations we set $M=61$ and fix the levels to
$\lambda_k = k$, $k = 0,1,2,\ldots,60$, $\lambda_{61} = \infty$.
We also perform subset simulations using a population $R$ varying between 1000 and 100000, and a fraction $p_0 = 0.1$ as seeds for the subsequent levels.

\begin{figure}
\centering
\includegraphics[width=0.5\textwidth]{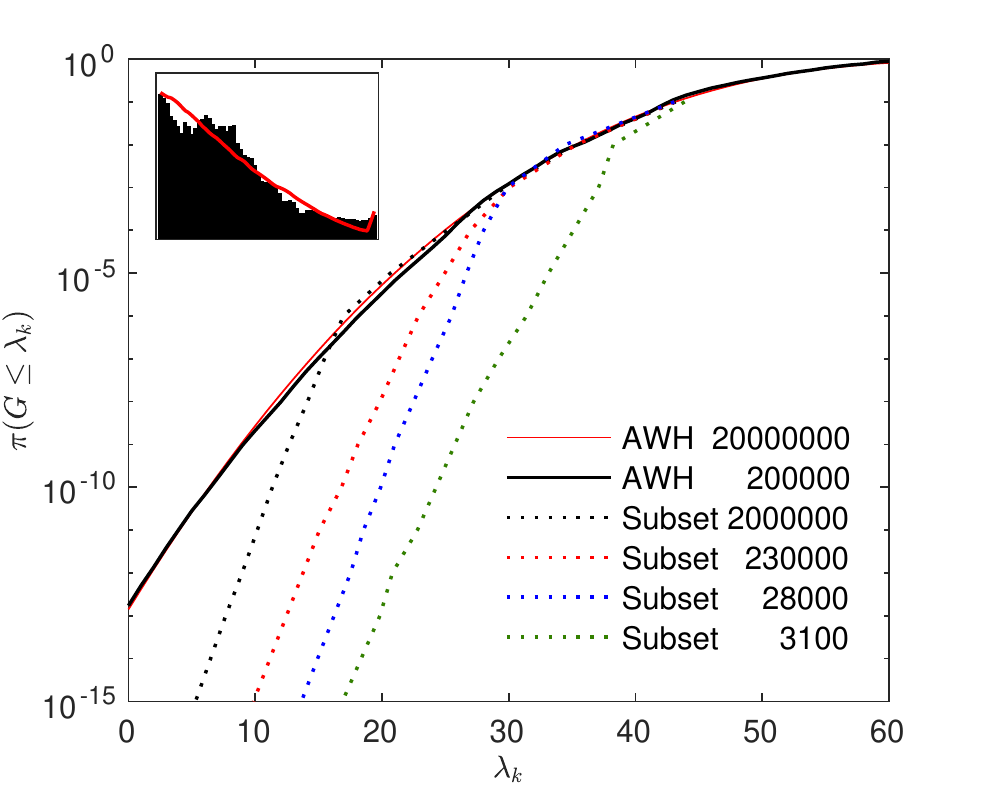}
\caption{\label{fig:FBM}Figure 3.
Failure probability for the FBM estimated using AWH and subset simulations.  The two AWH simulations are shown for
$2 \cdot 10^7$ and $2 \cdot 10^5$ iterations (with equally many function evaluations of $G(\cdot)$).
The corresponding weight histograms are shown in red and black, respectively, in the inset.
The subset simulations (dotted curves) used population sizes of $10^5, 10^4, 10^3$ and $10^2$.
The number of function evaluations are indicated in the figure.
}
\end{figure}

\begin{figure}
\centering
\includegraphics[width=0.5\textwidth]{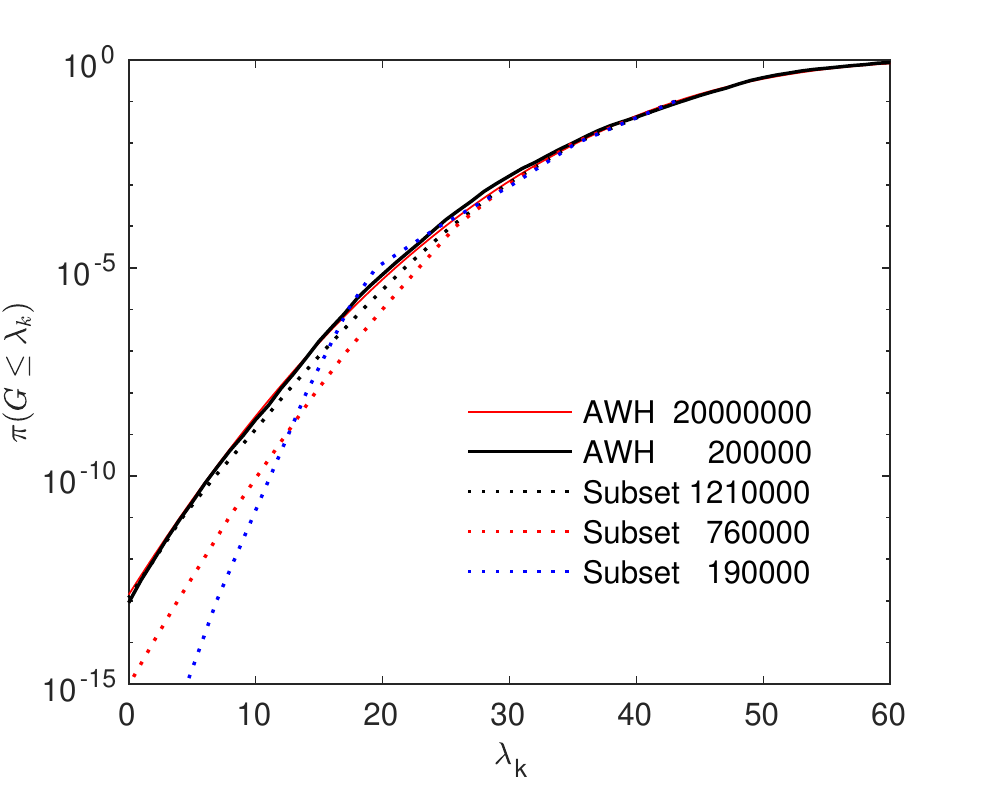}
\caption{\label{fig:FBM2}Figure 4.
The AWH simulations are as in Figure~\ref{fig:FBM},
while the subset simulation results shown here (dotted curves) are for a population size 10000, varying instead the number of MCMC updates at each level.
The lowest dotted curve used $1/p_0 = 10$ MCMC updates per level and seed,
the next used 50, and the final, which almost coincide with the AWH curve, used 100.
The number of function evaluations are indicated in the figure.
}
\end{figure}

In Figure~\ref{fig:FBM} we show the resulting failure probability $\pi(G \leq \lambda_k)$ as function of $\lambda$,
where $\lambda_0 = 0$ corresponds to failure for the given load $L$.
A very long AWH simulations using 20 million iterations is used to produce a nearly error free result (thin solid line), and an estimate of $\pi(\F) \approx 1.4 \cdot 10^{-13}$.
The thick solid line shows an AWH simulation using 200 000 iterations (equally many evaluations of $G(\cdot)$).
As can be seen the resulting curve quite accurately follows the previous one, and gave a failure probability of $\approx 1.7 \cdot 10^{-13}$.
The subset simulations on the other hand have difficulties for this model.
Subset simulations with population sizes 100, 1000, 10000, and 100000 were carried out, but none of them produced sensible results for the failure probability at $\lambda = 0$.
Instead the curves tend to bend down below some value of $\lambda$ which decrease with increasing population size.
Part of the difficulties can be explained by the slow dynamics of the MCMC moves used. If the number of MCMC sweeps are increased at each subset level the efficiency of the subset simulations improve. This is shown in Figure~\ref{fig:FBM2}.
By increasing the MCMC updates by a factor of 10 the AWH results are reproduced quite accurately.
Naturally, the number of function evaluations per level also increases by the same factor and reached over 1.2 million, which can be compared with the 200 000 used for the AWH simulation.

To asses the accuracy of the calculations we simulate a less extreme case with somewhat larger load $L=220$.
For reference, the failure probability was estimated to $4.8 \cdot 10^{-6}$ from a long AWH simulation of 50 million iterations.
We then repeated 50 shorter AWH simulations and 50 subset simulations.  The latter used a population of 10000, $p_0 = 0.1$, and 100 MCMC updates per seed and level.
The computational effort was the same in both cases, requiring 500000 evaluations of $G(\cdot)$ per simulation.
The RMS relative error from the AWH simulations was $\approx 0.25$, while it was $\approx 1.0$ from the subset simulations, i.e.\ around 4 times higher.

\section{Discussion and conclusions}

As demonstrated above the AWH simulation method can be effective for the estimation of small rare event probabilities, and in some cases surpass subset simulations.
Furthermore, the method is highly flexible and may for instance straightforwardly be generalized to include more than one limit state function in the same simulation.  It is also possible to simulate from the posterior probability given a set of observations in a Bayesian setting.
Another advantage is that it is rather easy to diagnose problems with the calculation by monitoring the weight histogram. If the normalized weight histogram deviates too much from the target distribution, this is a clear sign that the sampling is insufficient.
An example can be seen in Figure \ref{fig:W}, which shows how the weight histogram evolves towards the target distribution during the course of the simulation.

From a simulation point of view there is room for further optimization of the target distribution by allocating more samples in difficult regions~\citep{Lindahl2018}.
In the present study we have used a target distribution similar to the one employed in subset simulations in order to make fair comparisons. Although this seems to be a well performing generic choice, it deserves future study.

Compared to subset simulations, which move towards smaller and smaller regions of the configuration space, the AWH method is more sequential, going back and forth between the subset levels.
The large population of independent samples at the initial level of subset simulations is a great asset, that may give subset simulations an advantage -- if enough of that independence is carried through to the final level.
The AWH simulations on the other hand seem to handle well the situation where many MCMC steps are required to explore the relevant regions of the distribution.
For the normal distribution test case we found that subset simulations performed slightly better, although the AWH simulations were not far behind.
For the Fiber Bundle Model test case the situation was reversed.  This model, although highly idealized, contains many ingredients of more realistic structures.  Here AWH simulations generally performed well, whereas the subset simulations either failed to converge or produced less accurate results for the same computational effort.

A specific example on applications from the civil engineering field is reliability estimation of cables supporting large bridges. These cables are typically built-up by a bundle of high strength steel wires often modelled in analogy with the fiber bundle model presented herein, see e.g.~\citet{Faber2003}. The AWH method is also expected to perform well for more complex system reliability applications including correlated variables and multiple failure modes. This could include time dependent degradation of individual structural components, by e.g. fatigue or corrosion, in combination with the failure of the whole system. The latter remains to be tested.

\section*{Acknowledgements}

The presented research was funded and supported by the Rock Engineering Research Foundation (BeFo) [grant no. 424]. The research was conducted without involvement of the funding source.

\bibliography{rare}

\begin{thebibliography}{17}
\providecommand{\natexlab}[1]{#1}
\expandafter\ifx\csname urlstyle\endcsname\relax
  \providecommand{\doi}[1]{doi:\discretionary{}{}{}#1}\else
  \providecommand{\doi}{doi:\discretionary{}{}{}\begingroup
  \urlstyle{rm}\Url}\fi

\bibitem[{Au(2016)}]{Au2016}
Au, S.-K. (2016).
\newblock {On MCMC algorithm for Subset Simulation}.
\newblock \emph{Probabilistic Engineering Mechanics}, 43:117--120.
\newblock \doi{10.1016/j.probengmech.2015.12.003}.

\bibitem[{Au \& Beck(1999)}]{Au1999}
Au, S.-K. \& Beck, J.~L. (1999).
\newblock {A new adaptive importance sampling scheme for reliability
  calculations}.
\newblock \emph{Structural Safety}, 21(2):135--158.
\newblock \doi{10.1016/S0167-4730(99)00014-4}.

\bibitem[{Au \& Beck(2001)}]{Au2001}
Au, S.-K. \& Beck, J.~L. (2001).
\newblock {Estimation of small failure probabilities in high dimensions by
  subset simulation}.
\newblock \emph{Probabilistic Engineering Mechanics}, 16(4):263--277.
\newblock \doi{10.1016/S0266-8920(01)00019-4}.

\bibitem[{Au \& Wang(2014)}]{Au2014}
Au, S.-K. \& Wang, Y. (2014).
\newblock \emph{{Engineering Risk Assessment with Subset Simulation}}.
\newblock John Wiley {\&} Sons, Singapore.
\newblock \doi{10.1002/9781118398050}.

\bibitem[{Cohen et~al.(2009)Cohen, Lehmann, \& Or}]{Cohen2009}
Cohen, D., Lehmann, P., \& Or, D. (2009).
\newblock Fiber bundle model for multiscale modeling of hydromechanical
  triggering of shallow landslides.
\newblock \emph{Water Resources Research}, 45(10).
\newblock \doi{https://doi.org/10.1029/2009WR007889}.

\bibitem[{Daniels(1945)}]{Daniels1945}
Daniels, H. (1945).
\newblock {The statistical theory of the strength of bundles of threads. I}.
\newblock \emph{Proceedings of the Royal Society of London. Series A.
  Mathematical and Physical Sciences}, 183(995):405--435.
\newblock \doi{10.1098/rspa.1945.0011}.

\bibitem[{Faber et~al.(2003)Faber, Engelund, \& Rackwitz}]{Faber2003}
Faber, M., Engelund, S., \& Rackwitz, R. (2003).
\newblock Aspects of parallel wire cable reliability.
\newblock \emph{Structural Safety}, 25(2):201--225.
\newblock \doi{https://doi.org/10.1016/S0167-4730(02)00057-7}.

\bibitem[{Hasofer \& Lind(1974)}]{Hasofer1974}
Hasofer, A. \& Lind, N. (1974).
\newblock An exact and invariant first order reliability format.
\newblock \emph{Journal of Engineering Mechanics}, 100(1):111--121.

\bibitem[{Lidmar(2012)}]{AWH}
Lidmar, J. (2012).
\newblock {Improving the efficiency of extended ensemble simulations: The
  accelerated weight histogram method}.
\newblock \emph{Physical Review E}, 85(5):056708.
\newblock \doi{10.1103/PhysRevE.85.056708}.

\bibitem[{Lindahl et~al.(2018)Lindahl, Lidmar, \& Hess}]{Lindahl2018}
Lindahl, V., Lidmar, J., \& Hess, B. (2018).
\newblock {Riemann metric approach to optimal sampling of multidimensional
  free-energy landscapes}.
\newblock \emph{Physical Review E}, 98(2):023312.
\newblock \doi{10.1103/PhysRevE.98.023312}.

\bibitem[{Papaioannou et~al.(2015)Papaioannou, Betz, Zwirglmaier, \&
  Straub}]{Papaioannou2015}
Papaioannou, I., Betz, W., Zwirglmaier, K., \& Straub, D. (2015).
\newblock {MCMC algorithms for Subset Simulation}.
\newblock \emph{Probabilistic Engineering Mechanics}, 41(2):89--103.
\newblock \doi{10.1016/j.probengmech.2015.06.006}.

\bibitem[{Papaioannou et~al.(2016)Papaioannou, Papadimitriou, \&
  Straub}]{Papaioannou2016}
Papaioannou, I., Papadimitriou, C., \& Straub, D. (2016).
\newblock {Sequential importance sampling for structural reliability analysis}.
\newblock \emph{Structural Safety}, 62:66--75.
\newblock \doi{10.1016/j.strusafe.2016.06.002}.

\bibitem[{Peirce(1926)}]{Peirce1926}
Peirce, F.~T. (1926).
\newblock The weakest link theorems on the strength of long and of composite
  specimens.
\newblock \emph{Journal of the Textile Institute Transactions},
  17(7):T355--T368.
\newblock \doi{10.1080/19447027.1926.10599953}.

\bibitem[{Pradhan et~al.(2010)Pradhan, Hansen, \& Chakrabarti}]{Pradhan2010}
Pradhan, S., Hansen, A., \& Chakrabarti, B.~K. (2010).
\newblock {Failure processes in elastic fiber bundles}.
\newblock \emph{Reviews of Modern Physics}, 82(1):499--555.
\newblock \doi{10.1103/RevModPhys.82.499}.

\bibitem[{Rackwitz \& Flessler(1978)}]{Rackwitz1978}
Rackwitz, R. \& Flessler, B. (1978).
\newblock {Structural reliability under combined random load sequences}.
\newblock \emph{Computers {\&} Structures}, 9(5):489--494.
\newblock \doi{10.1016/0045-7949(78)90046-9}.

\bibitem[{Sornette(1992)}]{Sornette1992}
Sornette, D. (1992).
\newblock Mean-field solution of a block-spring model of earthquakes.
\newblock \emph{J. Phys. I France}, 2(11):2089--2096.
\newblock \doi{10.1051/jp1:1992269}.

\bibitem[{Straub et~al.(2016)Straub, Papaioannou, \& Betz}]{Straub2016}
Straub, D., Papaioannou, I., \& Betz, W. (2016).
\newblock {Bayesian analysis of rare events}.
\newblock \emph{Journal of Computational Physics}, 314:538--556.
\newblock \doi{10.1016/j.jcp.2016.03.018}.

\end{thebibliography}

\end{document}